\documentclass[sigconf]{acmart}

\AtBeginDocument{%
  }

 \setcopyright{rightsretained}
 \copyrightyear{2024}

\usepackage{framed}
\usepackage{array}
  \usepackage{graphicx}
  \graphicspath{{img/}}
  \DeclareGraphicsExtensions{.eps}
\usepackage[tight,footnotesize]{subfigure}

\usepackage{listings}

\usepackage{amsfonts}

\usepackage{xcolor}
\usepackage{mdframed}

\usepackage{todonotes}
\usepackage{sidecap}

\newcommand\YAMLcolonstyle{\color{red}\mdseries}
\newcommand\YAMLkeystyle{\color{black}\bfseries}
\newcommand\YAMLvaluestyle{\color{blue}\mdseries}

\makeatletter

\newcommand\language@yaml{yaml}

\expandafter\expandafter\expandafter\lstdefinelanguage
\expandafter{\language@yaml}
{
  keywords={true,false,null,y,n},
  keywordstyle=\color{darkgray}\bfseries,
  basicstyle=\YAMLkeystyle,                                 
  sensitive=false,
  comment=[l]{\#},
  morecomment=[s]{/*}{*/},
  commentstyle=\color{purple}\ttfamily,
  stringstyle=\YAMLvaluestyle\ttfamily,
  moredelim=[l][\color{orange}]{\&},
  moredelim=[l][\color{magenta}]{*},
  moredelim=**[il][\YAMLcolonstyle{:}\YAMLvaluestyle]{:},   
  morestring=[b]',
  morestring=[b]",
  literate =    {---}{{\ProcessThreeDashes}}3
                {>}{{\textcolor{red}\textgreater}}1     
                {|}{{\textcolor{red}\textbar}}1 
                {\ -\ }{{\mdseries\ -\ }}3,
}

\lst@AddToHook{EveryLine}{\ifx\lst@language\language@yaml\YAMLkeystyle\fi}
\makeatother

\newcommand\ProcessThreeDashes{\llap{\color{cyan}\mdseries-{-}-}}

\begin{document}

\title{Wilkins: HPC In Situ Workflows Made Easy}


\author{Orcun Yildiz}
\email{oyildiz@anl.gov}
\affiliation{%
  \institution{Argonne National Laboratory}
  \streetaddress{9700 S Cass Ave}
  \city{Lemont}
  \state{IL}
  \country{USA}
  \postcode{60439}
}

\author{Dmitriy Morozov}
\email{dmorozov@lbl.org}
\affiliation{%
  \institution{Lawrence Berkeley National Laboratory}
  \city{Berkeley}
    \state{CA}
  \country{USA}
}

\author{Arnur Nigmetov}
\email{anigmetov@lbl.org}
\affiliation{%
  \institution{Lawrence Berkeley National Laboratory}
  \city{Berkeley}
    \state{CA}
  \country{USA}
}
\author{Bogdan Nicolae}
\email{bnicolae@anl.gov}
\affiliation{%
  \institution{Argonne National Laboratory}
  \streetaddress{9700 S Cass Ave}
  \city{Lemont}
  \state{IL}
  \country{USA}
  \postcode{60439}
}
\author{Tom Peterka}
\email{tpeterka@mcs.anl.gov}
\affiliation{%
  \institution{Argonne National Laboratory}
  \streetaddress{9700 S Cass Ave}
  \city{Lemont}
  \state{IL}
  \country{USA}
  \postcode{60439}
}


\begin{abstract}
 In situ approaches can accelerate the pace of scientific discoveries by allowing scientists to perform data analysis at simulation time. Current in situ workflow systems, however, face challenges in handling the growing complexity and diverse computational requirements of scientific tasks. In this work, we present Wilkins, an in situ workflow system that is designed for ease-of-use while providing scalable and efficient execution of workflow tasks. Wilkins provides a flexible workflow description interface, employs a high-performance data transport layer based on HDF5, and supports tasks with disparate data rates by providing a flow control mechanism. Wilkins seamlessly couples scientific tasks that already use HDF5, without requiring task code modifications. We demonstrate the above features using both synthetic benchmarks and two science use cases in materials science and cosmology.
\end{abstract}



\maketitle

\section{Introduction}\label{sec:Introduction}

In situ workflows have gained traction in the high-performance computing (HPC) community because of the need to analyze increasing data volumes, together with the ever-growing 
gap between the computation and I/O capabilities of HPC systems. 
In situ workflows run within a single HPC system as a collection of multiple tasks, which are often large and parallel programs. These tasks communicate over memory or the interconnect of the HPC system, bypassing the parallel file system. Avoiding physical storage minimizes the I/O time and accelerates the pace of scientific discoveries. 

Despite their potential advantages, challenges for in situ workflows include the growing complexity and heterogeneity of today's scientific computing, which pose several problems that are addressed in this article. First, the workflow system should enable seamless coupling of user task codes, while providing a flexible interface to specify diverse data and computation requirements of these tasks. In particular, the workflow interface should support specification of today's complex workflows including computational steering and ensembles of tasks. Second, user tasks may employ a wide variety of data models. This heterogeneity of the data is even more evident with the growing number of AI tasks being incorporated in in situ workflows. The workflow system should provide a data model abstraction through which users can specify their view of data across heterogeneous tasks. Third, in situ workflows often include tasks with disparate data rates, requiring efficient flow control strategies to mitigate communication bottlenecks between tasks. 

Another key factor is the usability of in situ workflows. The workflow systems
should be easy to use while being able to express the different requirements of
users. One common concern among users is the amount of required modifications to
their task codes. Unfortunately, the current state of the art often requires
changes to user codes, where users manually need to insert workflow API calls
into their codes to be able to run them within the in situ workflow system. Such
code modifications can be cumbersome, depending on the level of such changes, and further impede adoption of workflow systems. Ideally, the same code should be able to run standalone as a single executable and as part of a workflow.

Driven by the needs of today's computational science campaigns, we introduce Wilkins, an in situ workflow system with the following features: 

\begin{itemize}
    \item Ease of adoption, providing scalable and efficient execution of workflow tasks without requiring any task code changes.
    \item A flexible workflow description interface that supports various workflow topologies ranging from simple linear workflows to complex ensembles.
    \item A high-performance data transport layer based on the rich HDF5 data model.
    \item A flow control mechanism to support efficient coupling of in situ workflow tasks with different rates.
\end{itemize}

We demonstrate the above features with both synthetic experiments and two different science use cases. The first is from materials science, where a workflow is developed for capturing a rare nucleation event. This requires orchestrating an ensemble of multiple molecular dynamics simulation instances coupled to a parallel in situ feature detector. In the second use case, the in situ workflow consists of a cosmological simulation code coupled to a parallel analysis task that identifies regions of high dark-matter density. These tasks have disparate computation rates, requiring efficient flow control strategies.

The remainder of this paper is organized as follows. Section~\ref{sec:Background} presents background and related work. Section~\ref{sec:Methodology} explains the design and implementation of Wilkins. Section~\ref{sec:Experimental Methodology} presents our  experimental results in both synthetic benchmarks as well as two representative science use cases. Section~\ref{sec:Conclusion} concludes the paper with a summary and a look toward the future.

\section{Background and Related Work}\label{sec:Background}

We first provide a brief background on in situ workflows. Then, we present the related work by categorizing in situ workflows according to their workflow description interfaces and data transfer mechanisms.

\subsection{In situ workflows}

Scientific computing encompasses various interconnected computational tasks. In situ workflow systems have been developed over the years by the HPC community to automate the dependencies and data exchanges between these tasks, eliminating the need for manual management. In situ workflows are designed to run within a single HPC system, launching all tasks concurrently. Data transfer between these tasks is done through memory or interconnect of the HPC system instead of the physical storage. Representative of such systems include ADIOS~\cite{boyuka2014transparent}, Damaris~\cite{dorier2016damaris}, Decaf~\cite{yildiz2022decaf}, ParaView Catalyst~\cite{ayachit2015paraview}, SENSEI~\cite{ayachit2016sensei}, and VisIt Libsim~\cite{kuhlen2011parallel}. 

\subsection{Workflow description interfaces}

Most in situ workflow systems use a static declarative interface in the form of a workflow configuration file to define the workflow. For instance, Decaf~\cite{yildiz2022decaf} and FlowVR~\cite{dreher2014flexible} workflow systems use a Python script for workflow graph description, while ADIOS~\cite{boyuka2014transparent}, Damaris~\cite{dorier2016damaris}, and VisIt Libsim~\cite{kuhlen2011parallel} all use an XML configuration file. Similarly, Wilkins provides a simple YAML configuration file for users to describe their workflows. Some workflow systems choose to employ an imperative interface. Henson~\cite{morozov2016master}, a cooperative multitasking system for in situ processing, follows this approach by having users directly modify the workflow master driver code. 

Alternatively, workflows can be defined implicitly using a programming language such as Swift/T~\cite{wozniak2013swift}, which schedules tasks according to data dependencies within the program. While Swift/T can handle complex workflows, users need to organize and compile their code into Swift modules.

One important aspect of workflow description interfaces is their extensibility while maintaining simplicity. In particular, the workflow interface should allow users to define complex scientific workflows with diverse requirements, ideally with minimal user effort. One example is ensemble workflows where there can be numerous workflow tasks and communication channels among them. There are some systems that are specifically designed for this type of workflows. Melissa~\cite{schouler2023melissa} is a framework to run large-scale ensembles and process them in situ. LibEnsemble~\cite{hudson2021libensemble} is a Python library that supports in situ processing of large-scale ensembles. DeepDriveMD~\cite{brace2022coupling} is a framework for ML-driven steering of molecular dynamics simulations that couples large-scale ensembles of AI and HPC tasks.
While we also support ensembles in Wilkins, our workflow description interface is generic and not specifically tailored to a particular category of workflows, such as ensembles.

\subsection{Data transfer mechanisms}

One key capability of workflows is to automate the data transfers between individual tasks within the workflow. Data transfer mechanisms vary among in situ workflows, but shared memory and network communication are the most common data transfer mechanisms. 

In in situ workflows where tasks are colocated on the same node, shared memory can offer benefits by enabling zero-copy communication. VisIt's Libsim and Paraview's Catalyst use shared-memory communication between analysis and visualization tasks, operating synchronously with the simulation within the same address space. Henson is another workflow system that supports shared-memory communication among colocated tasks on the same node. This is achieved by dynamically loading the executables of these tasks into the same address space.

When the workflow tasks are located on separate nodes within the same system, data can be transferred between the tasks using the system interconnect. This approach enables efficient parallel communication by eliminating the need for the parallel file system. Decaf~\cite{yildiz2022decaf} is a middleware for coupling parallel tasks in situ by establishing communication channels over HPC interconnects through MPI. Similarly, Damaris~\cite{dorier2016damaris} uses direct messaging via MPI between workflow tasks to exchange data. Wilkins also adopts this approach to provide efficient parallel communication between workflow tasks.

Some workflow systems opt to use a separate staging area when moving the data
between the tasks instead of direct messaging. This approach is often called
data staging; it requires extra resources for staging the data in an intermediate location. Systems such as DataSpaces~\cite{docan2012dataspaces}, FlexPath~\cite{dayal2014flexpath}, and Colza~\cite{dorier2022colza} adopt this approach, where they offload the data to a distributed memory space that is shared among multiple workflow tasks. Other approaches 
such as DataStates~\cite{DataStates20,DataStates-IPDPS22} retain multiple versions of datasets in the staging area, which enables the tasks to consume any past version of the dataset, not just the latest one.

While these in situ solutions offer efficient data transfers by avoiding physical storage, they share a common  requirement for modifications to task code. For instance, Decaf and DataSpaces both use a put/get API for data transfers which needs to be integrated into task codes. On the other hand, Wilkins does not require any changes to task codes if they already use HDF5 or one of the many front-ends to HDF5, such as HighFive~\cite{highfive22}, h5py~\cite{collette13}, NetCDF4~\cite{rew04}, SCORPIO~\cite{krishna20}, or Keras~\cite{gulli2017deep}.

\section{Design and Methodology}\label{sec:Methodology}

Wilkins is an in situ workflow system that enables heterogenous task specification and execution for in situ data processing. Wilkins provides a data-centric API for defining the workflow graph, creating and launching tasks, and establishing communicators between the tasks.

\subsection{Overall architecture}

\begin{figure}
    \centering
    \includegraphics[width=2.8in]{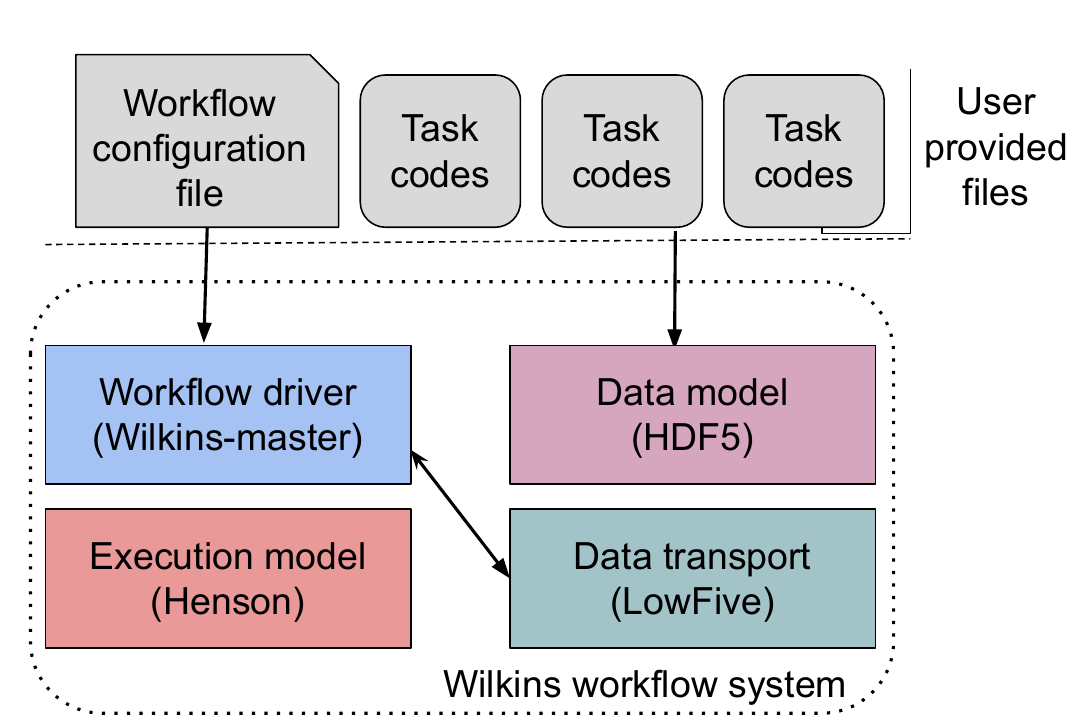}
    \caption{Overview of the Wilkins system.}
    \label{fig:wilkins}
\end{figure}

Figure~\ref{fig:wilkins} shows an overview of Wilkins and its main components, which are data transport, data model, workflow execution, and workflow driver. At its data transport layer, Wilkins uses the LowFive library~\cite{peterka2023lowfive}, which is a data model specification, redistribution, and communication library implemented as an HDF5 Virtual Object Layer (VOL) plugin. LowFive can be enabled either by setting environment variables or manually constructing a LowFive object, via the LowFive API, in the user task codes. Wilkins adopts the former approach to have task codes with no modifications. To execute the workflow tasks, Wilkins relies on Henson's execution model, where user task codes are compiled as shared objects~\cite{morozov2016master}. Besides shared objects, Henson uses coroutines as its main abstractions, which gives Wilkins extra flexibility when executing the tasks. At the workflow layer, Wilkins has a Python driver code, where all the workflow functions (e.g., data transfers, flow control) are defined through this code. This Python driver code is generic and provided by the Wilkins system, and users do not need to modify this code.

At the user level, users only need to provide the workflow configuration file and the constituent task codes. Linking the task codes as shared objects is often the only required additional step to use Wilkins.

\subsection{Data-centric workflow description}

Wilkins employs a data-centric workflow definition, where users indicate tasks' resource and data requirements using a workflow configuration file. Rather than specifying explicitly which tasks depend on others, users specify input and output data requirements in the form of file/dataset names.
By matching data requirements, Wilkins automatically creates the communication channels between the workflow tasks, and generates the workflow task graph as a representation of this workflow configuration file. Wilkins supports any directed-graph topology of tasks, including common patterns such as pipeline, fan-in, fan-out, ensembles of tasks, and cycles.

\begin{figure}[t]
    \centering
    \includegraphics[width=2.8in]{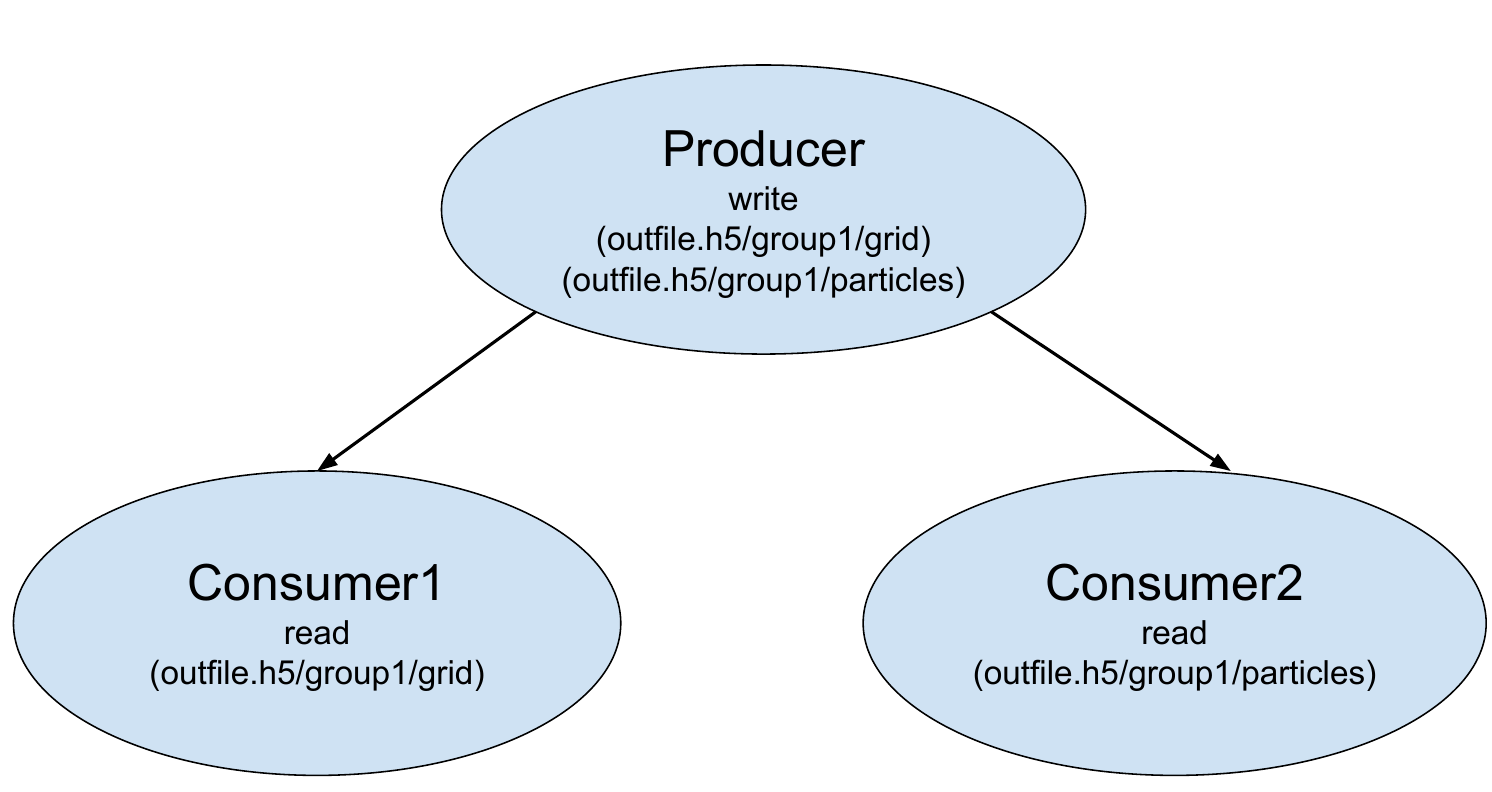}
    \caption{Example of three tasks coupled through Wilkins.}
    \label{fig:prod-2cons}
\end{figure}

Users describe their workflow definition in a YAML file. Listing~\ref{list:ymlfile} shows a sample YAML file representing a 3-task workflow consisting of 1 producer and 2 consumer tasks. The producer generates two different datasets---a structured grid of values and a list of particles---while the first and second consumer each require only the grid and particle datasets, respectively. Users describe these data requirements using the \textit{inport} and \textit{outport} fields in the YAML file. While the sample YAML file in Listing~\ref{list:ymlfile} uses full names for the file and dataset names, it is also possible to use matching patterns (e.g., \textit{*.h5/particles} can be used instead of \textit{outfile.h5/particles}).  Based on these requirements, Wilkins creates two communication channels: one channel between the producer and the first consumer for the grid dataset, and another channel between the producer and the second consumer for the particles dataset. In these channels, tasks will communicate using LowFive, Wilkins' data transport library, either through MPI or HDF5 files. Users can select the type of this communication in the YAML file by setting \textit{file} to 1 for using files or by setting \textit{memory} to 1 for using MPI. For instance, this example uses MPI in both of the communication channels between the coupled tasks. Figure~\ref{fig:prod-2cons} illustrates the workflow consisting of these three tasks coupled through Wilkins.

For the resource requirements of the tasks, users indicate the number of processes using the \textit{nprocs} field. Wilkins will assign these resources to the tasks and launch them accordingly. The execution model of Wilkins is described in Section~\ref{execModel}.

\begin{lstlisting}[float, captionpos=b, label=list:ymlfile, language=yaml, xleftmargin=1ex, basicstyle=\fontsize{7}{5}\tt, caption=Sample YAML file for describing a 3-task workflow consisting of 1 producer and 2 consumers.]
tasks: 
  - func: producer
    nprocs: 3
    outports:
      - filename: outfile.h5
        dsets:
          - name: /group1/grid
            file: 0
            memory: 1
          - name: /group1/particles
            file: 0
            memory: 1
  - func: consumer1
    nprocs: 5
    inports:
      - filename: outfile.h5
        dsets:
          - name: /group1/grid
            file: 0
            memory: 1
  - func: consumer2
    nprocs: 2
    inports:
      - filename: outfile.h5
        dsets:
          - name: /group1/particles
            file: 0
            memory: 1
\end{lstlisting}

\subsubsection{Defining ensembles}

Ensembles of tasks have become prevalent in scientific workflows. For instance, one common use case is to run the same simulation with different input parameters in hopes of capturing a rare scientific event~\cite{yildiz2019heterogeneous}. Other examples of ensembles arise in AI workflows performing hyperparameter optimization, or for uncertainty quantification~\cite{meyer2023high}. Such ensembles are often large-scale, requiring the orchestration of multiple concurrent tasks by the workflow system.

One question is how to specify an ensemble of tasks in a workflow configuration file. As there are often many tasks in an ensemble, we cannot expect users to list them explicitly. Instead, Wilkins provides an optional \textit{taskCount} field, where users can indicate the number of task instances in an ensemble. With this one extra field of information in the YAML file, Wilkins allows specification of various workflow graph topologies with ensembles of tasks including fan-in, fan-out, M to N, or combinations of those. Wilkins automatically creates the communication channels between the coupled ensemble tasks, without users having to explicitly list such dependencies thanks to its data-centric workflow description.  Listing~\ref{list:ensembles} shows a sample YAML file for describing ensembles with a fan-in topology, where four instances of a producer task are coupled to two instances of a consumer task. 

\begin{figure}[t]
    \centering
    \includegraphics[width=2in]{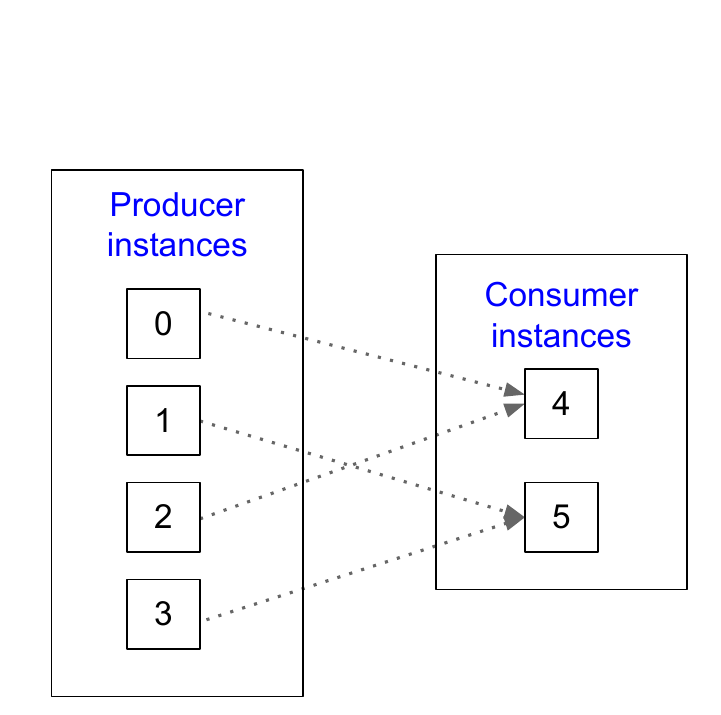}
    \caption{Example of ensemble coupling performed by Wilkins in a fan-in topology with 4 producer and 2 consumer instances.}
    \label{fig:ensembleCoupling}
\end{figure}

Figure~\ref{fig:ensembleCoupling} illustrates how Wilkins performs ensemble coupling of producer-consumer pairs in a fan-in topology with four producer and two consumer instances. 
For each matching data object, Wilkins creates a list of producer task indices and a list of consumer task indices. Wilkins then links these producer-consumer pairs by iterating through these indices in a round-robin fashion, as shown in Figure~\ref{fig:ensembleCoupling}. 

\begin{lstlisting}[float, captionpos=b, label=list:ensembles, language=yaml, xleftmargin=1ex, basicstyle=\fontsize{7}{5}\tt, caption=Sample YAML file for describing an ensemble of tasks with a fan-in topology.]
tasks: 
  - func: producer
    taskCount: 4 #Only change needed to define ensembles
    nprocs: 3
    outports:
      - filename: outfile.h5
        dsets:
          - name: /group1/grid
            file: 0
            memory: 1
  - func: consumer
    taskCount: 2 #Only change needed to define ensembles
    nprocs: 5
    inports:
      - filename: outfile.h5
        dsets:
          - name: /group1/grid
            file: 0
            memory: 1
\end{lstlisting}

\subsubsection{Defining subset of writers} Despite the advantages of parallel communication between the processes of workflow tasks, some simulations opt to perform serial I/O from a single process. For instance, the LAMMPS molecular dynamics simulation code first gathers all data to a single MPI process, and then this process writes the output serially~\cite{plimpton2007lammps}. 

To support such scenarios with serial or partially parallel writers, we introduce an optional $io\_proc$ field in the workflow configuration file. Users simply can indicate the number of writers in addition to the number of processes for the producer task. Then, Wilkins will assign this set of processes (starting from process 0) as I/O processes, while the remaining processes will only participate in the task execution (e.g., simulation) without performing any I/O operations.

This feature is implemented in the workflow driver code, which first checks whether a producer process is an I/O process based on the workflow configuration file. If so, the Wilkins driver creates a LowFive object and sets its properties (e.g., memory, file) in order for this process to participate in the data exchange. Local communicators and intercommunicators between the tasks provided to the LowFive object only involves I/O processes, and other processes do not participate in these communicator creation, which is handled by Wilkins. If the process is not an I/O process, the workflow driver simply skips all these I/O-related steps, and only executes this process as part of the producer task. 

\subsection{Workflow driver: Wilkins-master}

The Wilkins runtime, \textit{Wilkins-master}, is written in Python and serves as the main workflow driver to execute the workflow. \textit{Wilkins-master} orchestrates all the different functions within the workflow (e.g., launching tasks, data transfers, ensembles, flow control) as specified by the users in the workflow configuration file. Users do not need to modify the \textit{Wilkins-master} code to use any of the Wilkins capabilities.

\textit{Wilkins-master} first starts by reading the workflow configuration file to create the workflow graph. Based on this file, it creates local communicators for the tasks and intercommunicators between the interconnected tasks. Then, \textit{Wilkins-master} creates the LowFive plugin for the data transport layer. Next, it sets LowFive properties such as whether to perform data transfers using memory or files. After that, several Wilkins capabilities are defined, such as ensembles or flow control if they are specified in the configuration file. \textit{Wilkins-master} also checks whether there are any custom actions, which can be specific to particular use cases. We detail in Section~\ref{execModel} how users can specify custom actions through external Python scripts. Ultimately, \textit{Wilkins-master} launches the workflow.

\subsection{Data model and data transport library}

Wilkins employs the data model of LowFive library. HDF5~\cite{folk2011overview} is one of the most common data models, and as a VOL plugin, LowFive benefits from HDF5’s rich metadata describing the data model while affording users the familiarity of HDF5.

In its data transport layer, Wilkins uses the data redistribution components of LowFive, which enables data redistribution from M  to N processes. Wilkins allows coupled tasks to communicate both in situ using in-memory data and MPI message passing, and through traditional HDF5 files if needed. Users can select these different communication mechanisms via the workflow configuration file. 

We have extended the LowFive library by developing a callback functionality. In Wilkins, we use these callbacks to provide additional capabilities such as flow control. For example, we can exchange data between coupled tasks at a reduced frequency, rather than exchanging data at every iteration. Another scenario is custom callbacks, where users can define custom actions upon a specific I/O operation such as dataset open or file close. We will see examples of such callbacks in the next subsections.

\subsection{Execution model}
\label{execModel}

In a Wilkins workflow, user task codes can be serial or parallel; they can also have different languages such as C/C++, Python, or Fortran. Wilkins executes the user codes as a single-program-multiple-data (SPMD) application, thus having access to the MPI\_COMM\_WORLD across all ranks. Wilkins partitions this communicator and presents restricted MPI\_COMM\_WORLDs to the user codes, relying on Henson's PMPI tooling to make this transparent. This way the user codes see only their restricted world communicator, and the user codes are still written in a singular standalone fashion using this world communicator, as if they were its only users. Wilkins manages the partitioning of the global communicator into different local communicators, one for each task, as well as the intercommunicators connecting them. This process is entirely transparent to the users. Users only need to compile their codes as shared objects to execute them with Wilkins.

\subsubsection{Support for different consumer types}

In today's scientific workflows, we can categorize tasks into three types: i) producers such as HPC simulations that generate data periodically, ii) consumers such as analysis or visualization tasks that consume these data, and iii) intermediate tasks such as data processing that are both producers and consumers in a pipeline. Moreover, for the tasks that consume data, we can have two different types:

\paragraph{\textbf{Stateful consumer}} Such consumers maintain state information about the previous executions (e.g., timesteps). For instance, particle tracing codes need to keep information on the current trajectory of a particle~\cite{guo2017situ} while advecting the particle through the next step.

\paragraph{\textbf{Stateless consumer}} Such consumers do not maintain any information regarding their previous executions, as each run of a stateless task is entirely independent. For example, a feature detector code used in the analysis of molecular dynamics checks the number of nucleated atoms in each simulation timestep to determine whether nucleation is happening~\cite{yildiz2019heterogeneous}, with no relationship to previous timesteps.

Wilkins' execution model supports all these different task types, including both stateful and stateless consumers, transparent to the user. Wilkins first launches all these tasks as coroutines. Once the consumer tasks are completed, Wilkins uses a LowFive callback to query producers whether there are more data to consume. Producers respond to this query with the list of filenames that need to be consumed, or an empty list if no more data will be generated (all done). With this query logic, Wilkins handles both stateful and stateless consumer types. A stateful consumer is launched once and runs until completion for the number of timesteps or iterations as defined by the user. On the other hand, stateless consumers are launched as many times as there are incoming data to consume.

\subsubsection{Support for user-defined actions}

There can be scenarios that require custom workflow actions such as time- or data-dependent behaviors. For instance, users can request to transfer data between tasks only if the data value exceeds some predefined threshold. The \textit{Wilkins-master} code that executes the workflow is generic and does not support such custom actions by default. 

To support such custom actions, we explored two options: i) allowing users to modify the Wilkins master code directly (similar to the workflow systems with imperative interfaces such as Henson) or ii) letting users define these custom actions in an external Python script, which the Wilkins runtime incorporates. These options have tradeoffs with respect to usability and extensibility. While the first option, an imperative interface, provides more extensibility by exposing the workflow runtime to the user, it introduces additional complexity as users would need to be familiar with the \textit{Wilkins-master} code. We opted for a declarative interface, and decided that adopting the second option, defining external custom actions, would be a convenient middle ground between imperative and declarative interfaces. In our design, users define custom actions in a Python script using callbacks, and these callbacks allow imperative customization within an otherwise declarative interface.

Listing~\ref{list:sampleAction} shows a sample Python script representing the custom actions requested by the user. For instance, consider a scenario where the producer task performs two dataset write operations for the particles data including position and time values, but the analysis task only needs the position values of the particles dataset. Without modifying the producer task code, the user can provide this script to Wilkins, which will then perform data transfer between tasks after every second dataset write operation. In this script, user simply defines this custom action (\emph{custom\_cb}) in a callback at after dataset write (\emph{adw\_cb}), which delays the data transfer until the second occurrence of dataset write operation. We will see more examples of the use of callbacks in flow control and in the high-energy physics science use case.

\begin{lstlisting}[float,
                   caption={Sample custom action script that can be provided by the user.},
                   captionpos=b,
                   label=list:sampleAction,
                   language=python,
                   xleftmargin=1ex,
                   basicstyle=\fontsize{7}{5}\tt]

dw_counter = 0                   
def custom_cb(vol, rank):
    #after dataset write callback
    def adw_cb(s):
        global dw_counter
        dw_counter = dw_counter + 1
        if dw_counter % 2 == 0:
            #serving data at every 2 dataset write operation
            vol.serve_all(True, True)

    vol.set_after_dataset_write(adw_cb)
\end{lstlisting}

\subsection{Flow control}

In an in situ workflow, coupled tasks run concurrently, and wait for each other to send or receive data. Discrepancies among task throughputs can cause bottlenecks, where some tasks sit idle waiting for other tasks, resulting in wasted time and resources.
To alleviate such bottlenecks, Wilkins provides a flow control feature, where users can specify one of three different flow control strategies through the workflow configuration file:

\begin{itemize}
    \item \textbf{All}: This is the default strategy in Wilkins when users do not specify any flow control strategy. In this strategy, the producer waits until the consumer is ready to receive data. A slow consumer can result in idle time for the producer task.

    \item \textbf{Some}: With this strategy, users provide the desired frequency of the data exchange in order to accommodate a slow consumer. For instance, users can specify to consume data every N iterations, where N is equal to the desired frequency (e.g., 10 or 100). This strategy provides a tradeoff between blocking the producer and consuming at a lower frequency.  

    \item \textbf{Latest}: In this strategy, Wilkins drops older data in the communication channel and replaces them with the latest timestep from the producer once the consumer is ready. This strategy can be useful when the problem is time-critical, and scientists prefer to analyze the latest data points instead of older ones.
\end{itemize}

Specifying the flow control strategy requires adding only one extra field of information to the configuration file, \textit{io\_freq}, where users can set the above strategies by specifying $N > 1$ for the \emph{some} strategy, $1$ or $0$ for \emph{all}, and $-1$ for the \emph{latest} strategy.  

Wilkins enforces these different strategies for flow control using LowFive callbacks, transparent to the user. For instance, consider a simple workflow consisting of a faster producer coupled to a slower consumer, using the \textit{latest} flow control strategy.
In LowFive, the producer serves data to the consumer when the producer calls a file close operation.
When the \emph{latest} strategy is in place, Wilkins registers a callback before the file close function, where the producer checks whether there are any incoming requests from consumers before sending the data. If there are requests, then data transfer happens normally (the same as no flow control strategy). However, if there are no requests from the consumer, then the producer skips sending data at this timestep and proceeds with generating data for the next timestep. This process continues until the producer terminates. All these steps are part of LowFive and Wilkins, and are transparent to the user.

Such a flow control mechanism allows Wilkins to support heterogeneous workflows consisting of tasks with different data and computation rates.

\section{Experiments}\label{sec:Experimental Methodology}

Our experiments were conducted on the Bebop cluster at Argonne National Laboratory, which has 1,024 computing nodes. We employed nodes belonging to the Broadwell partition. The nodes in this partition are outfitted with 36-core Intel Xeon E5-2695v4 CPUs and 128 GB of DDR4 RAM. All nodes are connected to each other by an Intel Omni-Path interconnection network.

\subsection{Synthetic experiments}

We perform three different sets of experiments. For the first experiment, we use the hand-written code developed in ~\cite{peterka2023lowfive} to couple a producer and a consumer task that communicate using LowFive, without a workflow system on top. Then, we measure the overhead of Wilkins as a workflow system compared with that scenario. Second, we evaluate the flow control feature of Wilkins. Third, we demonstrate Wilkins' capability of supporting complex ensembles. 

\begin{table*}[t]
    \caption{Number of MPI processes for producer and consumer tasks and the total data size exchanged between them.
    }\label{tab:overhead}
       \begin{tabular}{| p{0.1\textwidth} | p{0.1\textwidth} | p{0.1\textwidth} | p{0.15\textwidth} | p{0.15\textwidth} | p{0.15\textwidth} }
       \textbf{Workflow size (procs)} & \textbf{Producer size (procs)} & \textbf{Consumer size (procs)}  & \textbf{Total data size ($10^6$/proc) (GiB)} & \textbf{Total data size ($10^7$/proc) (GiB)} & \textbf{Total data size ($10^8$/proc) (GiB)}\\
	   \hline
	   4  & 3 & 1 & 0.06 & 0.6 & 6\\ 
	   16  & 12 & 4 & 0.22 & 2.2 & 22\\ 
       64  & 48 & 16 & 0.99 & 9.9 & 99\\ 
       256  & 192 & 64 & 3.54 & 35.4 & 354\\ 
       1024  & 768 & 256 & 14.34 & 143.4 & 1434\\ 
   \end{tabular}
\end{table*}

For the synthetic benchmarks, we follow the approach used by Peterka et al.~\cite{peterka2023lowfive}. For the first two sets of experiments, we have a linear 2-node workflow coupling one producer and one consumer task. In the ensemble experiments, we vary the number of producer and consumer instances representing various workflow topologies. 

We generate synthetic data containing two datasets: one is a regular grid comprising 64-bit unsigned integer scalar values, and the other one is a list of particles, where each particle is a 3-d vector of 32-bit floating-point values. Per producer process, there are $10^6$ regularly structured grid points and $10^6$ particles. Each grid point and particle occupies 8 bytes and 12 bytes, respectively. Consequently, the total data per producer process is 19 MiB. We report the average times taken over 3 trials.

\subsubsection{Overhead of Wilkins compared with LowFive}

In this overhead experiment, we perform a weak scaling test by increasing the total data size proportionally with the number of producer processes. The producer generates the grid and particles datasets, and the consumer reads both of them. We allocate three-fourth of the processes to the producer, and the remaining one-fourth to the consumer task. For this overhead experiments, we also use larger data sizes with $10^7$ and $10^8$ grid points and particles per MPI process. Table~\ref{tab:overhead} shows the number of MPI processes for each task and total data sizes.

\begin{figure}
    \centering
    \includegraphics[width=2.8in]{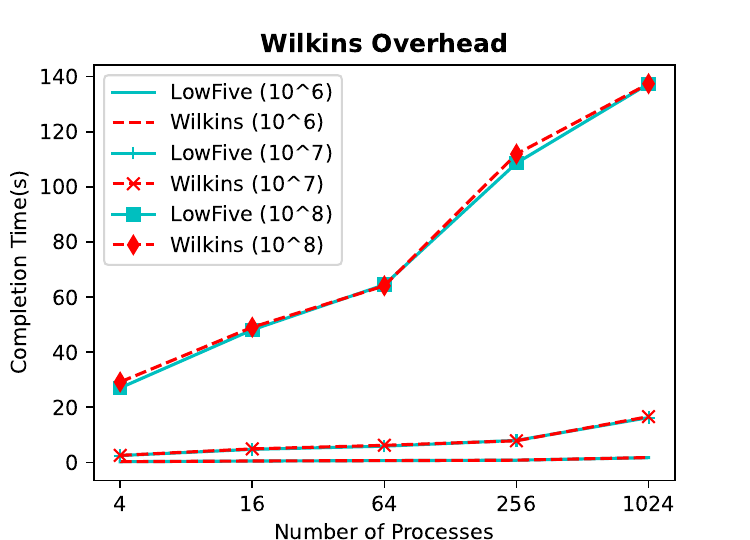} 
    \caption{Time to write/read grid and particles between 1 producer and 1 consumer task, comparing using LowFive alone with Wilkins.}
    \label{fig:overhead}
\end{figure}

Figure~\ref{fig:overhead} shows the time to write/read grid and particles datasets between the producer and consumer tasks in a weak scaling regime. As we can see from the results, the overhead of Wilkins is negligible for all data sizes. The difference between using LowFive standalone and with Wilkins at 1K processes is only 2\%.

\subsubsection{Flow control}

In these experiments, we use 512 processes for both producer and consumer tasks. We use the sleep function to emulate the computation behavior of tasks. For the producer task, we use 2 seconds sleep. For the consumer task, we emulate three different slow consumers as 2x, 5x, and 10x slow consumers by adding 4 seconds, 10 seconds, and 20 seconds sleep to the consumer tasks. The producer task runs for a total of 10 timesteps generating grid and particles datasets. We employ three different flow control strategies: i) \emph{all}---producer task serving  data at every timestep, ii) \emph{some}---producer task serving data at every 2, 5, or 10 timesteps, and iii) \emph{latest}---producer task serving data when the consumer signals that it is ready. For the \emph{some} strategy, we run with $N=2$ for the 2x slow consumer, $N=5$ for the 5x slow consumer, and $N=10$ for the 10x slow consumer. Table~\ref{tab:flowControl} shows the completion time of the workflow under these different strategies for each consumer task with different rates. We observe that using the \emph{some} and \emph{latest} flow control strategies bring up to 4.7x and 4.6x time savings, respectively. As expected, time savings are larger for the workflow with the slowest consumer (10x sleep). For instance, while the time savings are 1.6x with 2x slow consumer, this is 4.7x with the 10x slow consumer, with the \emph{some} flow control strategy. Time savings gained with the flow control strategies are due to the fact that the producer task does not have to wait for the slow consumer at every timestep, and can continue without serving to the next timestep when using the \emph{some} and \emph{latest} flow control strategies.

\begin{table*}[h]
    \caption{Completion time for the workflow coupling a producer and a (2x, 5x, and 10x) slow consumer under different flow control strategies. 
    }\label{tab:flowControl}
          \begin{tabular}{| p{0.1\textwidth} | p{0.2\textwidth} | p{0.2\textwidth} | p{0.2\textwidth}}
       \textbf{Strategy} & \textbf{Completion time (2x)} & \textbf{Completion time (5x)} & \textbf{Completion time (10x)}\\
	   \hline
	   All  & 51 seconds & 111.7 seconds & 211.7 seconds   \\ 
	   Some  & 31.2 seconds & 35 seconds & 44.9 seconds \\
       Latest  & 33.5 seconds & 38 seconds & 45.8 seconds\\
   \end{tabular}
\end{table*}

\begin{figure}
 \centering
 \subfigure[All]{
   \centering
   \includegraphics[width=0.5\textwidth]{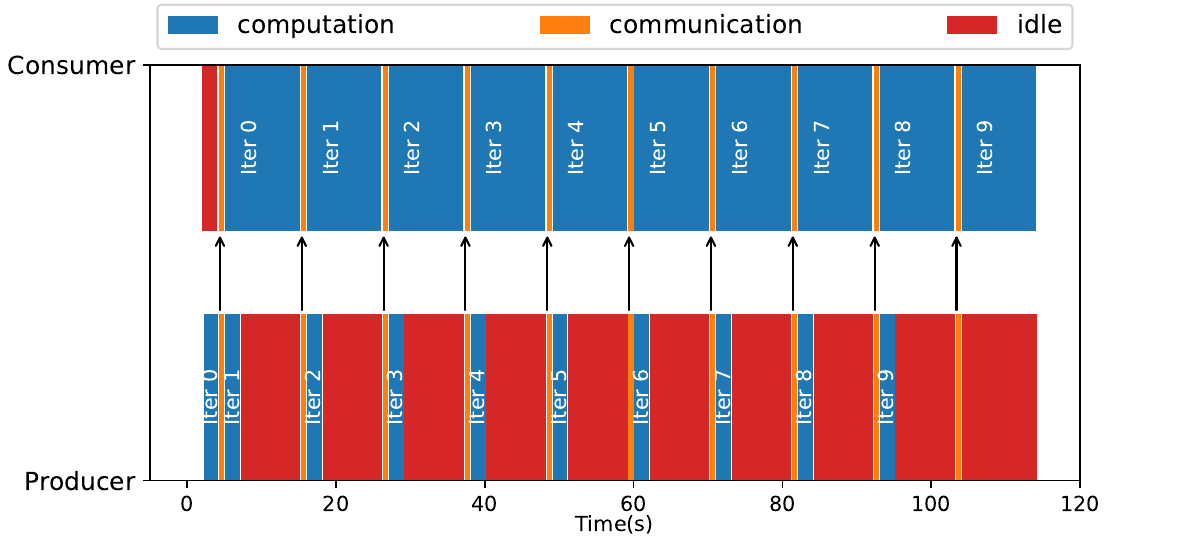} 
   \label{fig:fc-all}
 }
 \subfigure[Some]{
 \centering
 \includegraphics[width=0.5\textwidth]{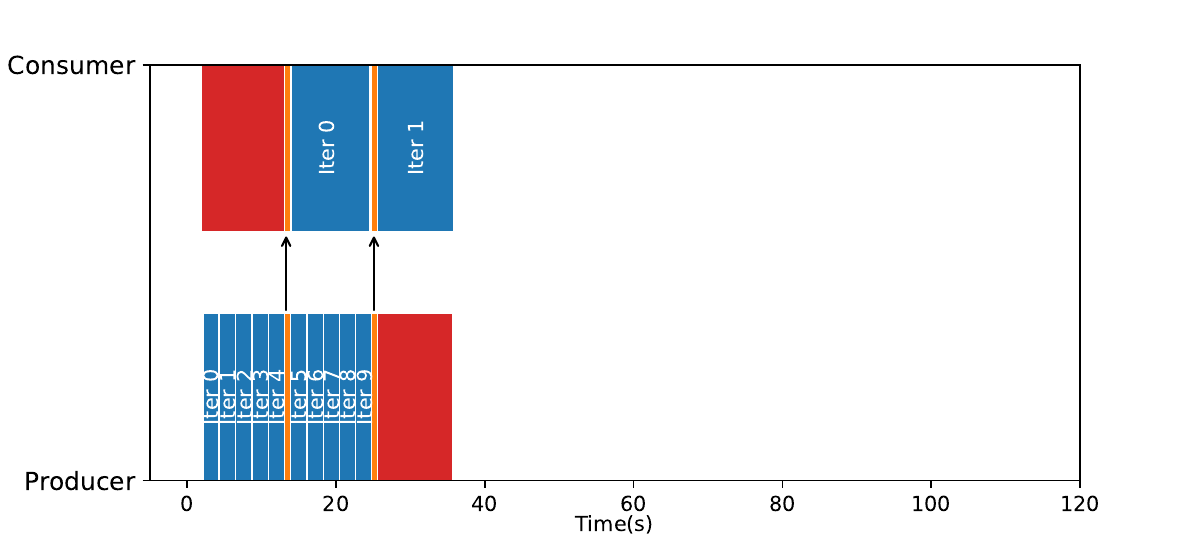}
 \label{fig:fc-some}
 }
  \subfigure[Latest]{
 \centering
 \includegraphics[width=0.5\textwidth]{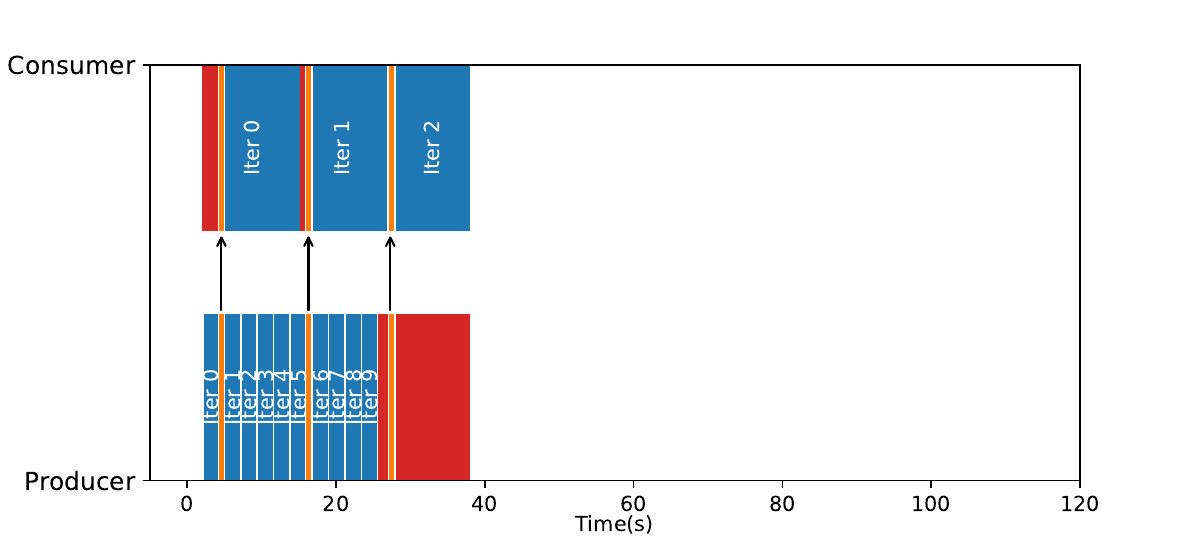}
 \label{fig:fc-latest}
 }
 \caption{Gantt charts for the execution of producer and 5x slow consumer for 10 iterations under different flow control strategies.} 
 \label{fig:fc-gantt}
\end{figure}

To further highlight the reduction in idle time for the producer, we illustrate the timeline for the execution of producer and 5x slow consumer under different flow control strategies in Figure~\ref{fig:fc-gantt}. Blue bars represent the computation, while red ones represent the idle time for workflow tasks. We show the data transfer between tasks with an arrow and orange bars. With \emph{all} flow control strategy, we can see that the producer task has to wait for the slow consumer at every timestep for the data transfer, resulting in significant idle time. In contrast, with the \emph{some} and \emph{latest} flow control strategies, these idle times are avoided where the producer task has to wait for the consumer only at the end of the workflow execution.

\subsubsection{Scaling of ensembles}

\begin{figure*}
    \centering
    \includegraphics[width=4in]{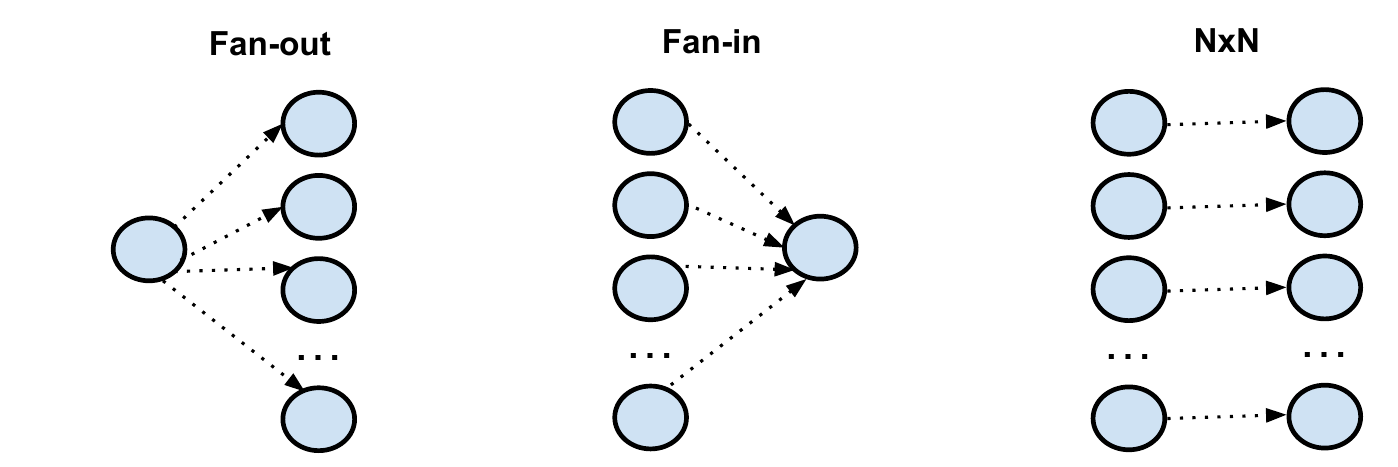}
    \caption{Example of three different ensemble topologies: fan-out, fan-in, and NxN.}
    \label{fig:ensembles}
\end{figure*}

In these ensemble experiments, we use 2 processes for both producer and consumer instances. We vary the number of these instances to represent three different ensemble topologies: fan-out, fan-in, and NxN. Examples of these topologies are shown in Figure~\ref{fig:ensembles}.

\begin{figure}
    \centering
    \includegraphics[width=2.8in]{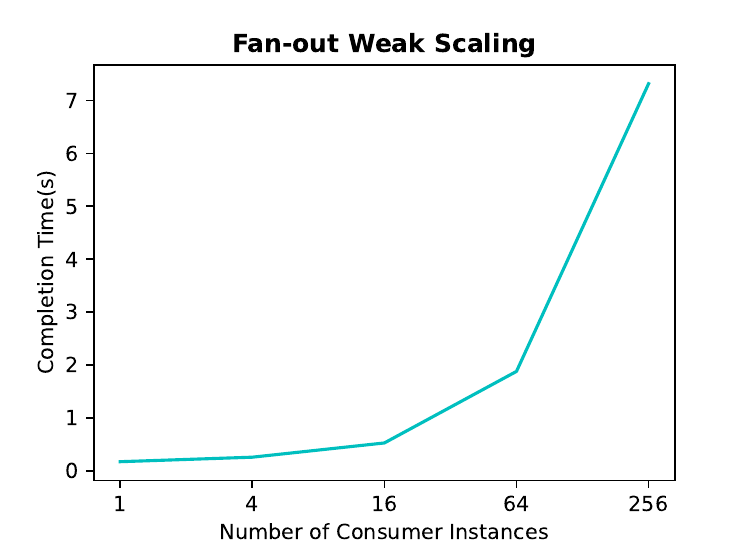} 
    \caption{Time to write/read the grid and particles between 1 producer and different numbers of consumer instances in a fan-out topology.}
    \label{fig:fanout}
\end{figure}

First, we analyze the time required to write/read the grid and particles between a single producer and different numbers of consumer instances in a fan-out topology. Figure~\ref{fig:fanout} shows the results, where we use 1, 4, 16, 64, and 256 consumer instances. We can see that total time increases almost linearly with the number of consumer instances. For example, while the completion time is around 0.6 seconds with 16 consumer instances, this time increases to 8.2 seconds with 256 consumer instances. This is due to the fact that the producer has to send the grid and particle datasets to each consumer instance, sequentially. 

\begin{figure}
    \centering
    \includegraphics[width=2.8in]{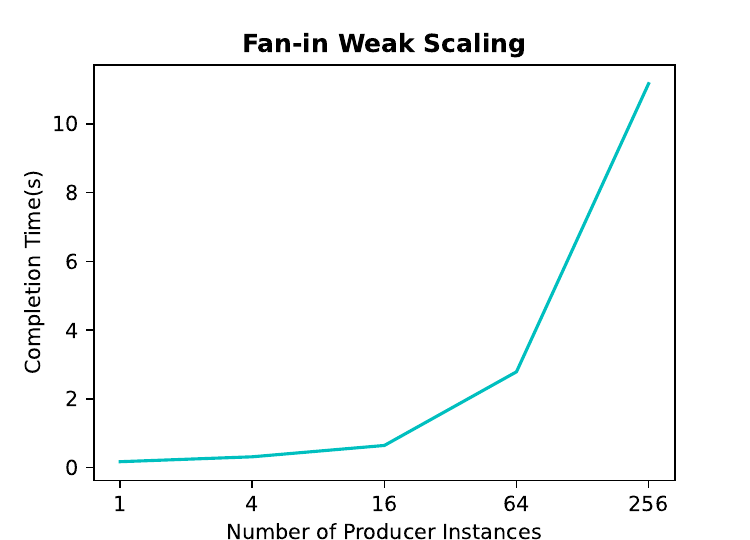}
    \caption{Time to write/read the grid and particles between different numbers of producer instances and a consumer in a fan-in topology.}
    \label{fig:fanin}
\end{figure}

Next, we evaluate Wilkins' support for the fan-in topology by varying the number of producer instances. Figure~\ref{fig:fanin} shows the results. Similarly to fan-out results, we see that total time increases almost linearly with the number of producer instances as the consumer has to read from each producer instance. 

\begin{figure}
    \centering
    \includegraphics[width=2.8in]{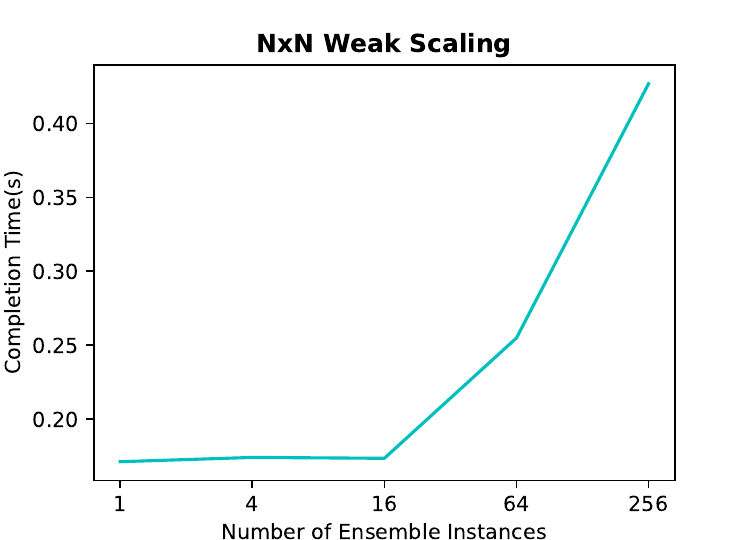}
    \caption{Time to write/read the grid and particles between different numbers of producer and consumer instances in an NxN topology.}
    \label{fig:NxN}
\end{figure}

Lastly, we evaluate the time required to write/read the grid and particles between different number of producer and consumer instances in an NxN topology. Figure~\ref{fig:NxN} displays the results, where we use 1, 4, 16, 64, and 256 instances for both producer and consumer tasks. Unlike the fan-out and fan-in topologies, we can observe that the time difference is only minimal when using different numbers of ensemble instances. This different behavior is expected as we have a one-to-one relationship between producer and consumer instances in an NxN topology. The slight increase in the total time can be attributed to the increased network contention at larger scales. 

\subsection{Science use cases}

\subsubsection{Materials science}

Nucleation occurs as a material cools and crystallizes, e.g., when water freezes. Understanding nucleation in material systems is important for better understanding of several natural and technological systems~\cite{chan2019machine, greer2016overview, gettelman2012climate}. Nucleation, however, is a stochastic event that requires a large number of molecules to reveal its kinetics. Simulating nucleation is difficult, especially in the initial phases of simulation when only a few atoms have crystallized.

One way scientists simulate nucleation is to run many instances of small simulations, requiring an ensemble of tasks where simulations with different initial configurations are coupled to analysis tasks. In this workflow, we couple a LAMMPS molecular dynamics simulation~\cite{plimpton2007lammps} in situ with a parallel feature detector that finds crystals in a diamond-shaped lattice~\cite{yildiz2019heterogeneous}. To create the ensemble, we use N instances for both simulation and analysis tasks in an NxN topology. To define these ensemble tasks, we only need to add the \textit{taskCount} information to the workflow configuration file. Listing~\ref{list:lammps-detector} shows the configuration file for this molecular dynamics workflow with 64 ensemble instances.

\begin{lstlisting}[float, captionpos=b, label=list:lammps-detector, language=yaml, xleftmargin=1ex, basicstyle=\fontsize{7}{5}\tt, caption=Sample YAML file for describing the molecular dynamics workflow for simulating nucleation with 64 ensemble instances.]
tasks: 
  - func: freeze
    taskCount: 64 #Only change needed to define ensembles
    nprocs: 32
    nwriters: 1 #Only rank 0 performs I/O
    outports:
      - filename: dump-h5md.h5
        dsets:
          - name: /particles/*
            file: 0
            memory: 1
  - func: detector
    taskCount: 64 #Only change needed to define ensembles
    nprocs: 8
    inports:
      - filename: dump-h5md.h5
        dsets:
          - name: /particles/*
            file: 0
            memory: 1
\end{lstlisting}

In LAMMPS's I/O scheme, all simulation data are gathered to rank 0 before they are written serially. This undermines Wilkins' capacity for efficient parallel communications. On the other hand, this demonstrates the applicability of the subset writers feature of Wilkins, where we only need to set the number of writers (i.e., $io\_proc$) to 1 in the configuration file, as shown in Listing~\ref{list:lammps-detector}. Furthermore, LAMMPS supports writing HDF5 files, therefore, no modifications are needed to execute LAMMPS with Wilkins; we only had to compile LAMMPS as a shared library with HDF5 support. 

\begin{figure}
    \centering
    \includegraphics[width=2.8in]{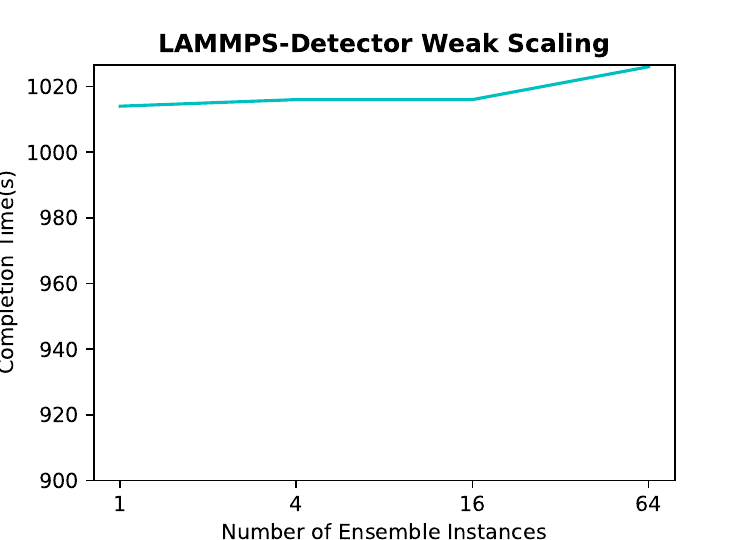}
    \caption{Completion time of the molecular dynamics workflow consisting of LAMMPS coupled to a diamond detector with different number of ensemble instances.}
    \label{fig:lammps-detector}
\end{figure}

In these experiments, we use 32 processes for each LAMMPS instance, and 8 processes for each analysis task. We run LAMMPS for 1,000,000 time steps with a water model composed of 4,360 atoms, and we perform the diamond structure analysis every 10,000 iterations. To conduct this experiment, we vary the number of ensemble instances from 1 up to 64. Figure~\ref{fig:lammps-detector} shows the completion time under these scenarios. The results demonstrate that Wilkins can support execution of different number of ensemble instances without adding any significant overhead, in particular when there are a matching number of consumer instances in an NxN configuration. For example, the difference in completion time between a single instance and 64 ensemble instances is only 1.2\%. 

In terms of ease-of-use, no changes were made to the simulation or the feature detector source code to execute inside Wilkins, and to launch multiple instances in an ensemble, only one line was added to the producer and consumer task descriptions in the YAML workflow configuration file.

\subsubsection{High-energy physics} 

The second use case is motivated by cosmology; in particular, halo finding in simulations of dark matter. The in situ workflow consists of Nyx~\cite{almgren2013nyx}, a parallel cosmological simulation code, coupled to a smaller-scale parallel analysis task called Reeber~\cite{friesen2016situ, nigmetov2019local} that identifies high regions of density, called halos, at certain time steps.

AMReX~\cite{zhang2019amrex}, a framework designed for massively parallel adaptive mesh refinement computations, serves as the PDE solver of Nyx simulation code, as well as providing I/O, writing the simulation data into a single HDF5 file. Reeber supports reading HDF5. As these user codes already use HDF5, no modifications were needed to execute them with Wilkins.

Ideally, a code utilizing parallel I/O would perform the following sequence of I/O operations. It would collectively create or open a file once from all MPI processes. This would be followed by some number of I/O operations, in parallel from all MPI processes. Eventually the file would be closed, again collectively from all MPI processes. LowFive is designed for this pattern, initiating the serving of data from a producer task to a consumer task upon the producer closing the file and the consumer opening the file. This assumes that the file close and file open occur exactly once, from all MPI processes in the task, as described above.

However, not all simulation codes perform I/O in this way, and Nyx is not the only code that violates this pattern. For various reasons---often related to poor I/O performance when accessing small amounts of data collectively---Nyx and other codes often employ patterns where a single MPI process creates or opens a file, performs small I/O operations from that single process, closes the file, and then all MPI processes re-open the file collectively for bulk data access in parallel. The file is opened and closed twice, the first time by a single MPI process, and the second time by all the processes in the task.

Such custom I/O patterns, which vary from one code to another, break the assumptions in LowFive and Wilkins about when and how to serve data from producer to consumer. Fortunately, there is an elegant solution to incorporating custom I/O actions such as above. We added to the LowFive library custom callback functions at various execution points such as before and after file open and close. The user can program custom actions into those functions, e.g., counting the number of times a file is closed and delaying serving data until the second occurrence. In Wilkins, those custom functions are implemented by the user in a separate Python script, so that the user task code is unaffected.

Listing~\ref{list:customActions} shows the custom functions used in this cosmology use case, where there are two custom callback functions at after file close (\emph{afc\_cb}) and before file open (\emph{bfo\_cb}). In the after file close callback, process 0 broadcasts the data to other processes at the first file close, and serves data to the consumer at the second time, while other processes serve data upon (the one and only) file close. In the before file open callback, all processes other than 0 receive the data from process 0.

\begin{lstlisting}[float,
                   caption={User action script provided by the user for enforcing the custom HDF5 I/O mechanism of Nyx.},
                   captionpos=b,
                   label=list:customActions,
                   language=python,
                   xleftmargin=1ex,
                   basicstyle=\fontsize{7}{5}\tt]

                   
def nyx(vol, rank):
    #after file close callback
    def afc_callback(s):
        if rank != 0:
            #other ranks, serving data
            vol.serve_all(True, True)
            vol.clear_files()
        else:
            if vol.file_close_counter % 2 == 0:
                #rank 0, serving data
                vol.serve_all(True, True)
                vol.clear_files()
            else:
                #rank 0 broadcasting files to other ranks
                vol.broadcast_files()

    #before file open callback
    def bfo_cb(s):
        if rank != 0:
            #other ranks receiving files from rank 0
            vol.broadcast_files()
    #setting the callbacks in the VOL plugin
    vol.set_after_file_close(afc_callback)
    vol.set_before_file_open(bfo_cb)
\end{lstlisting}

The user provides this script for custom actions and indicates it in the YAML file by setting the optional actions field with the name of this script file and the defined custom user function (i.e., $actions: ["actions", "nyx"]$). Listing~\ref{list:nyx-reeber} shows the configuration file for this cosmology workflow.

\begin{lstlisting}[float, captionpos=b, label=list:nyx-reeber, language=yaml, xleftmargin=1ex, basicstyle=\fontsize{7}{5}\tt, caption=Sample YAML file for describing the cosmology workflow.]
tasks: 
  - func: nyx
    nprocs: 1024
    actions: ["actions", "nyx"]
    outports:
      - filename: plt*.h5
        dsets:
          - name: /level_0/density
            file: 0
            memory: 1
  - func: reeber
    nprocs: 64
    inports:
      - filename: plt*.h5
        io_freq: 2 #Setting the some flow control strategy
        dsets:
          - name: /level_0/density
            file: 0
            memory: 1
\end{lstlisting}

Depending on the timestep, number of MPI processes, number of dark matter particles, number of halos, and density cutoff threshold, Reeber can take longer to analyze a timestep than Nyx takes to compute it. To prevent idling of Nyx and wasting computational resources while waiting for Reeber, we make use of the flow control strategies in Wilkins.

In these experiments, we use 1024 processes for Nyx and 64 processes for Reeber. The Nyx simulation has a grid size of $256^3$, and it produces 20 snapshots to be analyzed by Reeber. For this experiment we intentionally slowed Reeber down even further by computing the halos a number of times (i.e., 100), making the effect of flow control readily apparent. This allowed us to run Nyx with a smaller number of processes and for a shorter period of time, saving computing resources. 
We employ two different flow control strategies: i) \emph{all}: Nyx serving  data at every timestep and ii) \emph{some}: Nyx serving data at every $n$ timesteps, in this case we vary $n$ as $n=2$, $n=5$, and $n=10$. Table~\ref{tab:flowControl-nyx-reeber} shows the completion time of the workflow under these different strategies. Similarly to the synthetic experiments, we observe that using the \emph{some} flow control strategy brings up to 7.7x time savings compared with the \emph{all} strategy. 

In terms of ease-of-use, we added one \emph{actions} line and one \emph{io\_freq} line to the vanilla YAML configuration file in order to take advantage of custom callbacks and flow control, and made no changes to Nyx or Reeber source code in order to work with Wilkins. The only other required user file is the action script, which is a short Python code consisting of less than 25 lines.

\begin{table}[h]
    \caption{Completion time for the cosmology workflow coupling Nyx and Reeber task codes under different flow control strategies.}
    \label{tab:flowControl-nyx-reeber}
          \begin{tabular}{| p{0.15\textwidth} | p{0.15\textwidth} }
       \textbf{Strategy} & \textbf{Completion time} \\
	   \hline
	   All  & 5,421 seconds   \\ 
	   Some ($n=2$)  & 2,754 seconds \\
       Some ($n=5$)  & 1,084 seconds \\
       Some ($n=10$)  & 702 seconds \\
   \end{tabular}
\end{table}


\section{Conclusion}\label{sec:Conclusion}

We have introduced Wilkins, an in situ workflow system designed with ease-of-use in mind for addressing the needs of today's scientific campaigns. Wilkins has a flexible data-centric workflow interface that supports the definition of several workflow topologies ranging from simple linear workflows to complex ensembles. Wilkins provides efficient communication of scientific tasks through LowFive, a high-performance data transport layer based on the rich HDF5 data model. Wilkins also allows users to define custom I/O actions through callbacks to meet different requirements of scientific tasks. Wilkins provides a flow control mechanism to manage tasks with different data rates. We used both synthetic benchmarks and two representative science use cases in materials science and cosmology to evaluate these features. The results demonstrated that Wilkins can support complex scientific workflows with diverse requirements while requiring no task code modifications.

Several avenues remain open for future work. Currently, Wilkins uses a static workflow configuration file, and cannot respond to dynamic changes in the requirements of scientific tasks during execution. We are currently working on extending Wilkins to support dynamic workflow changes. We are also collaborating closely with domain scientists to engage Wilkins in more science use cases. In particular, we are exploring use cases that couple HPC and AI applications, which can further demonstrate the usability of Wilkins in heterogeneous workflows.

\begin{acks}
This material is based upon work supported by the U.S. Department of Energy, Office 
of Science, Office of Advanced Scientific Computing Research, under contract numbers DE-AC02-06CH11357, and DE-AC02-05CH11231, program manager Margaret Lentz.
We gratefully acknowledge the computing resources provided on Bebop, a high-performance computing cluster operated by the Laboratory Computing Resource Center at Argonne National Laboratory.
\end{acks}

\bibliographystyle{ACM-Reference-Format}
\bibliography{biblio}


\begin{thebibliography}{35}


\ifx \showCODEN    \undefined \def \showCODEN     #1{\unskip}     \fi
\ifx \showDOI      \undefined \def \showDOI       #1{#1}\fi
\ifx \showISBNx    \undefined \def \showISBNx     #1{\unskip}     \fi
\ifx \showISBNxiii \undefined \def \showISBNxiii  #1{\unskip}     \fi
\ifx \showISSN     \undefined \def \showISSN      #1{\unskip}     \fi
\ifx \showLCCN     \undefined \def \showLCCN      #1{\unskip}     \fi
\ifx \shownote     \undefined \def \shownote      #1{#1}          \fi
\ifx \showarticletitle \undefined \def \showarticletitle #1{#1}   \fi
\ifx \showURL      \undefined \def \showURL       {\relax}        \fi
\providecommand\bibfield[2]{#2}
\providecommand\bibinfo[2]{#2}
\providecommand\natexlab[1]{#1}
\providecommand\showeprint[2][]{arXiv:#2}

\bibitem[Almgren et~al\mbox{.}(2013)]%
        {almgren2013nyx}
\bibfield{author}{\bibinfo{person}{Ann~S Almgren}, \bibinfo{person}{John~B Bell}, \bibinfo{person}{Mike~J Lijewski}, \bibinfo{person}{Zarija Luki{\'c}}, {and} \bibinfo{person}{Ethan Van~Andel}.} \bibinfo{year}{2013}\natexlab{}.
\newblock \showarticletitle{Nyx: A massively parallel amr code for computational cosmology}.
\newblock \bibinfo{journal}{\emph{The Astrophysical Journal}} \bibinfo{volume}{765}, \bibinfo{number}{1} (\bibinfo{year}{2013}), \bibinfo{pages}{39}.
\newblock


\bibitem[Ayachit et~al\mbox{.}(2015)]%
        {ayachit2015paraview}
\bibfield{author}{\bibinfo{person}{Utkarsh Ayachit}, \bibinfo{person}{Andrew Bauer}, \bibinfo{person}{Berk Geveci}, \bibinfo{person}{Patrick O'Leary}, \bibinfo{person}{Kenneth Moreland}, \bibinfo{person}{Nathan Fabian}, {and} \bibinfo{person}{Jeffrey Mauldin}.} \bibinfo{year}{2015}\natexlab{}.
\newblock \showarticletitle{{ParaView} {C}atalyst: Enabling in situ data analysis and visualization}. In \bibinfo{booktitle}{\emph{Proceedings of the First Workshop on In Situ Infrastructures for Enabling Extreme-Scale Analysis and Visualization}}. ACM, \bibinfo{pages}{25--29}.
\newblock


\bibitem[Ayachit et~al\mbox{.}(2016)]%
        {ayachit2016sensei}
\bibfield{author}{\bibinfo{person}{Utkarsh Ayachit}, \bibinfo{person}{Brad Whitlock}, \bibinfo{person}{Matthew Wolf}, \bibinfo{person}{Burlen Loring}, \bibinfo{person}{Berk Geveci}, \bibinfo{person}{David Lonie}, {and} \bibinfo{person}{E Bethel}.} \bibinfo{year}{2016}\natexlab{}.
\newblock \showarticletitle{The {SENSEI} generic in situ interface}. In \bibinfo{booktitle}{\emph{Proceedings of the 2nd Workshop on In Situ Infrastructures for Enabling Extreme-scale Analysis and Visualization}}. IEEE Press, \bibinfo{pages}{40--44}.
\newblock


\bibitem[BlueBrain(2022)]%
        {highfive22}
\bibfield{author}{\bibinfo{person}{BlueBrain}.} \bibinfo{year}{2022}\natexlab{}.
\newblock \bibinfo{title}{{HighFive - HDF5 header-only C++ Library}}.
\newblock \bibinfo{howpublished}{\url{https://github.com/BlueBrain/HighFive}}.
\newblock


\bibitem[Boyuka et~al\mbox{.}(2014)]%
        {boyuka2014transparent}
\bibfield{author}{\bibinfo{person}{David~A Boyuka}, \bibinfo{person}{Sriram Lakshminarasimham}, \bibinfo{person}{Xiaocheng Zou}, \bibinfo{person}{Zhenhuan Gong}, \bibinfo{person}{John Jenkins}, \bibinfo{person}{Eric~R Schendel}, \bibinfo{person}{Norbert Podhorszki}, \bibinfo{person}{Qing Liu}, \bibinfo{person}{Scott Klasky}, {and} \bibinfo{person}{Nagiza~F Samatova}.} \bibinfo{year}{2014}\natexlab{}.
\newblock \showarticletitle{Transparent in situ data transformations in adios}. In \bibinfo{booktitle}{\emph{2014 14th IEEE/ACM International Symposium on Cluster, Cloud and Grid Computing}}. IEEE, \bibinfo{pages}{256--266}.
\newblock


\bibitem[Brace et~al\mbox{.}(2022)]%
        {brace2022coupling}
\bibfield{author}{\bibinfo{person}{Alexander Brace}, \bibinfo{person}{Igor Yakushin}, \bibinfo{person}{Heng Ma}, \bibinfo{person}{Anda Trifan}, \bibinfo{person}{Todd Munson}, \bibinfo{person}{Ian Foster}, \bibinfo{person}{Arvind Ramanathan}, \bibinfo{person}{Hyungro Lee}, \bibinfo{person}{Matteo Turilli}, {and} \bibinfo{person}{Shantenu Jha}.} \bibinfo{year}{2022}\natexlab{}.
\newblock \showarticletitle{Coupling streaming ai and hpc ensembles to achieve 100--1000$\times$ faster biomolecular simulations}. In \bibinfo{booktitle}{\emph{2022 IEEE International Parallel and Distributed Processing Symposium (IPDPS)}}. IEEE, \bibinfo{pages}{806--816}.
\newblock


\bibitem[Chan et~al\mbox{.}(2019)]%
        {chan2019machine}
\bibfield{author}{\bibinfo{person}{Henry Chan}, \bibinfo{person}{Mathew~J Cherukara}, \bibinfo{person}{Badri Narayanan}, \bibinfo{person}{Troy~D Loeffler}, \bibinfo{person}{Chris Benmore}, \bibinfo{person}{Stephen~K Gray}, {and} \bibinfo{person}{Subramanian~KRS Sankaranarayanan}.} \bibinfo{year}{2019}\natexlab{}.
\newblock \showarticletitle{Machine learning coarse grained models for water}.
\newblock \bibinfo{journal}{\emph{Nature communications}} \bibinfo{volume}{10}, \bibinfo{number}{1} (\bibinfo{year}{2019}), \bibinfo{pages}{379}.
\newblock


\bibitem[Collette(2013)]%
        {collette13}
\bibfield{author}{\bibinfo{person}{Andrew Collette}.} \bibinfo{year}{2013}\natexlab{}.
\newblock \bibinfo{booktitle}{\emph{{Python and HDF5: Unlocking Scientific Data}}}.
\newblock \bibinfo{publisher}{" O'Reilly Media, Inc."}.
\newblock


\bibitem[Dayal et~al\mbox{.}(2014)]%
        {dayal2014flexpath}
\bibfield{author}{\bibinfo{person}{Jai Dayal}, \bibinfo{person}{Drew Bratcher}, \bibinfo{person}{Greg Eisenhauer}, \bibinfo{person}{Karsten Schwan}, \bibinfo{person}{Matthew Wolf}, \bibinfo{person}{Xuechen Zhang}, \bibinfo{person}{Hasan Abbasi}, \bibinfo{person}{Scott Klasky}, {and} \bibinfo{person}{Norbert Podhorszki}.} \bibinfo{year}{2014}\natexlab{}.
\newblock \showarticletitle{Flexpath: Type-based publish/subscribe system for large-scale science analytics}. In \bibinfo{booktitle}{\emph{2014 14th IEEE/ACM International Symposium on Cluster, Cloud and Grid Computing}}. IEEE, \bibinfo{pages}{246--255}.
\newblock


\bibitem[Docan et~al\mbox{.}(2012)]%
        {docan2012dataspaces}
\bibfield{author}{\bibinfo{person}{Ciprian Docan}, \bibinfo{person}{Manish Parashar}, {and} \bibinfo{person}{Scott Klasky}.} \bibinfo{year}{2012}\natexlab{}.
\newblock \showarticletitle{Dataspaces: an interaction and coordination framework for coupled simulation workflows}.
\newblock \bibinfo{journal}{\emph{Cluster Computing}} \bibinfo{volume}{15}, \bibinfo{number}{2} (\bibinfo{year}{2012}), \bibinfo{pages}{163--181}.
\newblock


\bibitem[Dorier et~al\mbox{.}(2016)]%
        {dorier2016damaris}
\bibfield{author}{\bibinfo{person}{Matthieu Dorier}, \bibinfo{person}{Gabriel Antoniu}, \bibinfo{person}{Franck Cappello}, \bibinfo{person}{Marc Snir}, \bibinfo{person}{Robert Sisneros}, \bibinfo{person}{Orcun Yildiz}, \bibinfo{person}{Shadi Ibrahim}, \bibinfo{person}{Tom Peterka}, {and} \bibinfo{person}{Leigh Orf}.} \bibinfo{year}{2016}\natexlab{}.
\newblock \showarticletitle{Damaris: Addressing performance variability in data management for post-petascale simulations}.
\newblock \bibinfo{journal}{\emph{ACM Transactions on Parallel Computing (TOPC)}} \bibinfo{volume}{3}, \bibinfo{number}{3} (\bibinfo{year}{2016}), \bibinfo{pages}{15}.
\newblock


\bibitem[Dorier et~al\mbox{.}(2022)]%
        {dorier2022colza}
\bibfield{author}{\bibinfo{person}{Matthieu Dorier}, \bibinfo{person}{Zhe Wang}, \bibinfo{person}{Utkarsh Ayachit}, \bibinfo{person}{Shane Snyder}, \bibinfo{person}{Rob Ross}, {and} \bibinfo{person}{Manish Parashar}.} \bibinfo{year}{2022}\natexlab{}.
\newblock \showarticletitle{Colza: Enabling elastic in situ visualization for high-performance computing simulations}. In \bibinfo{booktitle}{\emph{2022 IEEE International Parallel and Distributed Processing Symposium (IPDPS)}}. IEEE, \bibinfo{pages}{538--548}.
\newblock


\bibitem[Dreher and Raffin(2014)]%
        {dreher2014flexible}
\bibfield{author}{\bibinfo{person}{Matthieu Dreher} {and} \bibinfo{person}{Bruno Raffin}.} \bibinfo{year}{2014}\natexlab{}.
\newblock \showarticletitle{A flexible framework for asynchronous in situ and in transit analytics for scientific simulations}. In \bibinfo{booktitle}{\emph{2014 14th IEEE/ACM International Symposium on Cluster, Cloud and Grid Computing}}. IEEE, \bibinfo{pages}{277--286}.
\newblock


\bibitem[Folk et~al\mbox{.}(2011)]%
        {folk2011overview}
\bibfield{author}{\bibinfo{person}{Mike Folk}, \bibinfo{person}{Gerd Heber}, \bibinfo{person}{Quincey Koziol}, \bibinfo{person}{Elena Pourmal}, {and} \bibinfo{person}{Dana Robinson}.} \bibinfo{year}{2011}\natexlab{}.
\newblock \showarticletitle{An overview of the HDF5 technology suite and its applications}. In \bibinfo{booktitle}{\emph{Proceedings of the EDBT/ICDT 2011 workshop on array databases}}. \bibinfo{pages}{36--47}.
\newblock


\bibitem[Friesen et~al\mbox{.}(2016)]%
        {friesen2016situ}
\bibfield{author}{\bibinfo{person}{Brian Friesen}, \bibinfo{person}{Ann Almgren}, \bibinfo{person}{Zarija Luki{\'c}}, \bibinfo{person}{Gunther Weber}, \bibinfo{person}{Dmitriy Morozov}, \bibinfo{person}{Vincent Beckner}, {and} \bibinfo{person}{Marcus Day}.} \bibinfo{year}{2016}\natexlab{}.
\newblock \showarticletitle{In situ and in-transit analysis of cosmological simulations}.
\newblock \bibinfo{journal}{\emph{Computational astrophysics and cosmology}} \bibinfo{volume}{3}, \bibinfo{number}{1} (\bibinfo{year}{2016}), \bibinfo{pages}{1--18}.
\newblock


\bibitem[Gettelman et~al\mbox{.}(2012)]%
        {gettelman2012climate}
\bibfield{author}{\bibinfo{person}{Andrew Gettelman}, \bibinfo{person}{Xiaohong Liu}, \bibinfo{person}{Donifan Barahona}, \bibinfo{person}{Ulrike Lohmann}, {and} \bibinfo{person}{Celia Chen}.} \bibinfo{year}{2012}\natexlab{}.
\newblock \showarticletitle{Climate impacts of ice nucleation}.
\newblock \bibinfo{journal}{\emph{Journal of geophysical research: Atmospheres}} \bibinfo{volume}{117}, \bibinfo{number}{D20} (\bibinfo{year}{2012}).
\newblock


\bibitem[Greer(2016)]%
        {greer2016overview}
\bibfield{author}{\bibinfo{person}{AL Greer}.} \bibinfo{year}{2016}\natexlab{}.
\newblock \showarticletitle{Overview: Application of heterogeneous nucleation in grain-refining of metals}.
\newblock \bibinfo{journal}{\emph{The Journal of chemical physics}} \bibinfo{volume}{145}, \bibinfo{number}{21} (\bibinfo{year}{2016}).
\newblock


\bibitem[Gulli and Pal(2017)]%
        {gulli2017deep}
\bibfield{author}{\bibinfo{person}{Antonio Gulli} {and} \bibinfo{person}{Sujit Pal}.} \bibinfo{year}{2017}\natexlab{}.
\newblock \bibinfo{booktitle}{\emph{Deep learning with Keras}}.
\newblock \bibinfo{publisher}{Packt Publishing Ltd}.
\newblock


\bibitem[Guo et~al\mbox{.}(2017)]%
        {guo2017situ}
\bibfield{author}{\bibinfo{person}{Hanqi Guo}, \bibinfo{person}{Tom Peterka}, {and} \bibinfo{person}{Andreas Glatz}.} \bibinfo{year}{2017}\natexlab{}.
\newblock \showarticletitle{In situ magnetic flux vortex visualization in time-dependent Ginzburg-Landau superconductor simulations}. In \bibinfo{booktitle}{\emph{2017 IEEE Pacific Visualization Symposium (PacificVis)}}. IEEE, \bibinfo{pages}{71--80}.
\newblock


\bibitem[Hudson et~al\mbox{.}(2021)]%
        {hudson2021libensemble}
\bibfield{author}{\bibinfo{person}{Stephen Hudson}, \bibinfo{person}{Jeffrey Larson}, \bibinfo{person}{John-Luke Navarro}, {and} \bibinfo{person}{Stefan~M Wild}.} \bibinfo{year}{2021}\natexlab{}.
\newblock \showarticletitle{libEnsemble: A library to coordinate the concurrent evaluation of dynamic ensembles of calculations}.
\newblock \bibinfo{journal}{\emph{IEEE Transactions on Parallel and Distributed Systems}} \bibinfo{volume}{33}, \bibinfo{number}{4} (\bibinfo{year}{2021}), \bibinfo{pages}{977--988}.
\newblock


\bibitem[Krishna(2020)]%
        {krishna20}
\bibfield{author}{\bibinfo{person}{Jayesh Krishna}.} \bibinfo{year}{2020}\natexlab{}.
\newblock \bibinfo{title}{Scorpio – Parallel I/O Library}.
\newblock \bibinfo{howpublished}{\url{https://e3sm.org/scorpio-parallel-io-library/}}.
\newblock


\bibitem[Kuhlen et~al\mbox{.}(2011)]%
        {kuhlen2011parallel}
\bibfield{author}{\bibinfo{person}{T Kuhlen}, \bibinfo{person}{R Pajarola}, {and} \bibinfo{person}{K Zhou}.} \bibinfo{year}{2011}\natexlab{}.
\newblock \showarticletitle{Parallel in situ coupling of simulation with a fully featured visualization system}. In \bibinfo{booktitle}{\emph{Proceedings of the 11th Eurographics Conference on Parallel Graphics and Visualization (EGPGV)}}.
\newblock


\bibitem[Meyer et~al\mbox{.}(2023)]%
        {meyer2023high}
\bibfield{author}{\bibinfo{person}{Lucas Meyer}, \bibinfo{person}{Marc Schouler}, \bibinfo{person}{Robert~Alexander Caulk}, \bibinfo{person}{Alejandro Rib{\'e}s}, {and} \bibinfo{person}{Bruno Raffin}.} \bibinfo{year}{2023}\natexlab{}.
\newblock \showarticletitle{High Throughput Training of Deep Surrogates from Large Ensemble Runs}. In \bibinfo{booktitle}{\emph{SC 2023-The International Conference for High Performance Computing, Networking, Storage, and Analysis}}. ACM, \bibinfo{pages}{1--14}.
\newblock


\bibitem[Morozov and Lukic(2016)]%
        {morozov2016master}
\bibfield{author}{\bibinfo{person}{Dmitriy Morozov} {and} \bibinfo{person}{Zarija Lukic}.} \bibinfo{year}{2016}\natexlab{}.
\newblock \showarticletitle{Master of puppets: {C}ooperative multitasking for in situ processing}. In \bibinfo{booktitle}{\emph{Proceedings of the 25th ACM International Symposium on High-Performance Parallel and Distributed Computing}}. ACM, \bibinfo{pages}{285--288}.
\newblock


\bibitem[Nicolae(2020)]%
        {DataStates20}
\bibfield{author}{\bibinfo{person}{Bogdan Nicolae}.} \bibinfo{year}{2020}\natexlab{}.
\newblock \showarticletitle{{DataStates: Towards Lightweight Data Models for Deep Learning}}. In \bibinfo{booktitle}{\emph{{SMC'20: The 2020 Smoky Mountains Computational Sciences and Engineering Conference}}}. \bibinfo{address}{Nashville, United States}, \bibinfo{pages}{117--129}.
\newblock
\urldef\tempurl%
\url{https://doi.org/10.1007/978-3-030-63393-6_8}
\showDOI{\tempurl}


\bibitem[Nicolae(2022)]%
        {DataStates-IPDPS22}
\bibfield{author}{\bibinfo{person}{Bogdan Nicolae}.} \bibinfo{year}{2022}\natexlab{}.
\newblock \showarticletitle{{Scalable Multi-Versioning Ordered Key-Value Stores with Persistent Memory Support}}. In \bibinfo{booktitle}{\emph{{IPDPS 2022: The 36th IEEE International Parallel and Distributed Processing Symposium}}}. \bibinfo{address}{Lyon, France}, \bibinfo{pages}{93--103}.
\newblock
\urldef\tempurl%
\url{https://doi.org/10.1109/IPDPS53621.2022.00018}
\showDOI{\tempurl}


\bibitem[Nigmetov and Morozov(2019)]%
        {nigmetov2019local}
\bibfield{author}{\bibinfo{person}{Arnur Nigmetov} {and} \bibinfo{person}{Dmitriy Morozov}.} \bibinfo{year}{2019}\natexlab{}.
\newblock \showarticletitle{Local-global merge tree computation with local exchanges}. In \bibinfo{booktitle}{\emph{Proceedings of the International Conference for High Performance Computing, Networking, Storage and Analysis}}. \bibinfo{pages}{1--13}.
\newblock


\bibitem[Peterka et~al\mbox{.}(2023)]%
        {peterka2023lowfive}
\bibfield{author}{\bibinfo{person}{Tom Peterka}, \bibinfo{person}{Dmitriy Morozov}, \bibinfo{person}{Arnur Nigmetov}, \bibinfo{person}{Orcun Yildiz}, \bibinfo{person}{Bogdan Nicolae}, {and} \bibinfo{person}{Philip~E Davis}.} \bibinfo{year}{2023}\natexlab{}.
\newblock \showarticletitle{LowFive: In Situ Data Transport for High-Performance Workflows}. In \bibinfo{booktitle}{\emph{IPDPS'23: The 37th IEEE International Parallel and Distributed Processing Symposium}}.
\newblock


\bibitem[Plimpton et~al\mbox{.}(2007)]%
        {plimpton2007lammps}
\bibfield{author}{\bibinfo{person}{Steve Plimpton}, \bibinfo{person}{Paul Crozier}, {and} \bibinfo{person}{Aidan Thompson}.} \bibinfo{year}{2007}\natexlab{}.
\newblock \showarticletitle{LAMMPS-large-scale atomic/molecular massively parallel simulator}.
\newblock \bibinfo{journal}{\emph{Sandia National Laboratories}}  \bibinfo{volume}{18} (\bibinfo{year}{2007}), \bibinfo{pages}{43}.
\newblock


\bibitem[Rew et~al\mbox{.}(2004)]%
        {rew04}
\bibfield{author}{\bibinfo{person}{Russell~K Rew}, \bibinfo{person}{B Ucar}, {and} \bibinfo{person}{EJ Hartnett}.} \bibinfo{year}{2004}\natexlab{}.
\newblock \showarticletitle{{Merging NetCDF and HDF5}}. In \bibinfo{booktitle}{\emph{20th Int. Conf. on Interactive Information and Processing Systems}}.
\newblock


\bibitem[Schouler et~al\mbox{.}(2023)]%
        {schouler2023melissa}
\bibfield{author}{\bibinfo{person}{Marc Schouler}, \bibinfo{person}{Robert~Alexander Caulk}, \bibinfo{person}{Lucas Meyer}, \bibinfo{person}{Th{\'e}ophile Terraz}, \bibinfo{person}{Christoph Conrads}, \bibinfo{person}{Sebastian Friedemann}, \bibinfo{person}{Achal Agarwal}, \bibinfo{person}{Juan~Manuel Baldonado}, \bibinfo{person}{Bart{\l}omiej Pogodzi{\'n}ski}, \bibinfo{person}{Anna Seku{\l}a}, {et~al\mbox{.}}} \bibinfo{year}{2023}\natexlab{}.
\newblock \showarticletitle{Melissa: coordinating large-scale ensemble runs for deep learning and sensitivity analyses}.
\newblock \bibinfo{journal}{\emph{Journal of Open Source Software}} \bibinfo{volume}{8}, \bibinfo{number}{86} (\bibinfo{year}{2023}), \bibinfo{pages}{5291}.
\newblock


\bibitem[Wozniak et~al\mbox{.}(2013)]%
        {wozniak2013swift}
\bibfield{author}{\bibinfo{person}{Justin~M Wozniak}, \bibinfo{person}{Timothy~G Armstrong}, \bibinfo{person}{Michael Wilde}, \bibinfo{person}{Daniel~S Katz}, \bibinfo{person}{Ewing Lusk}, {and} \bibinfo{person}{Ian~T Foster}.} \bibinfo{year}{2013}\natexlab{}.
\newblock \showarticletitle{Swift/t: Large-scale application composition via distributed-memory dataflow processing}. In \bibinfo{booktitle}{\emph{2013 13th IEEE/ACM International Symposium on Cluster, Cloud, and Grid Computing}}. IEEE, \bibinfo{pages}{95--102}.
\newblock


\bibitem[Yildiz et~al\mbox{.}(2022)]%
        {yildiz2022decaf}
\bibfield{author}{\bibinfo{person}{Orcun Yildiz}, \bibinfo{person}{Matthieu Dreher}, {and} \bibinfo{person}{Tom Peterka}.} \bibinfo{year}{2022}\natexlab{}.
\newblock \showarticletitle{Decaf: Decoupled Dataflows for In Situ Workflows}.
\newblock In \bibinfo{booktitle}{\emph{In Situ Visualization for Computational Science}}. \bibinfo{publisher}{Springer}, \bibinfo{pages}{137--158}.
\newblock


\bibitem[Yildiz et~al\mbox{.}(2019)]%
        {yildiz2019heterogeneous}
\bibfield{author}{\bibinfo{person}{Orcun Yildiz}, \bibinfo{person}{Jorge Ejarque}, \bibinfo{person}{Henry Chan}, \bibinfo{person}{Subramanian Sankaranarayanan}, \bibinfo{person}{Rosa~M Badia}, {and} \bibinfo{person}{Tom Peterka}.} \bibinfo{year}{2019}\natexlab{}.
\newblock \showarticletitle{Heterogeneous hierarchical workflow composition}.
\newblock \bibinfo{journal}{\emph{Computing in Science \& Engineering}} (\bibinfo{year}{2019}).
\newblock


\bibitem[Zhang et~al\mbox{.}(2019)]%
        {zhang2019amrex}
\bibfield{author}{\bibinfo{person}{Weiqun Zhang}, \bibinfo{person}{Ann Almgren}, \bibinfo{person}{Vince Beckner}, \bibinfo{person}{John Bell}, \bibinfo{person}{Johannes Blaschke}, \bibinfo{person}{Cy Chan}, \bibinfo{person}{Marcus Day}, \bibinfo{person}{Brian Friesen}, \bibinfo{person}{Kevin Gott}, \bibinfo{person}{Daniel Graves}, {et~al\mbox{.}}} \bibinfo{year}{2019}\natexlab{}.
\newblock \showarticletitle{AMReX: a framework for block-structured adaptive mesh refinement}.
\newblock \bibinfo{journal}{\emph{The Journal of Open Source Software}} \bibinfo{volume}{4}, \bibinfo{number}{37} (\bibinfo{year}{2019}), \bibinfo{pages}{1370}.
\newblock


\end{thebibliography}










\end{document}